\documentclass[prd,twocolumn,groupedaddress,nofootinbib,english,nobalancelastpage,floatfix,superscriptaddress,preprintnumbers]{revtex4-1}
\usepackage{graphicx}
\usepackage{amsmath,bm}
\usepackage[T1]{fontenc}
\usepackage{rotating}
\usepackage{float}
\usepackage{array}
\usepackage{multirow}	
\usepackage{placeins}	
\floatstyle{plaintop}

\usepackage{color}
\definecolor{naviBlue}{RGB}{0,0,128}
\usepackage[colorlinks  = true,
            linkcolor   = naviBlue,
            urlcolor    = naviBlue,
            citecolor   = naviBlue,
            anchorcolor = naviBlue]{hyperref}
\newcommand{\secref}[1]{\hyperref[sec::#1]{SECTION~\ref*{sec::#1}}}
\newcommand{\subsecref}[1]{\hyperref[subsec::#1]{SECTION.~\ref*{subsec::#1}}}
\newcommand{\figref}[1]{\hyperref[fig::#1]{FIG.$\,$\ref*{fig::#1}}}
\newcommand{\tabref}[1]{\hyperref[tab::#1]{TABLE$\,$\ref*{tab::#1}}}
\newcommand{\eqnref}[1]{\hyperref[eqn::#1]{Eq.$\,$(\ref*{eqn::#1})}}

\newcommand{\diff}{\mathrm{d}}

\newcommand{\msbot}{m_{\widetilde b}}
\newcommand{\sbot}{\widetilde b}
\newcommand{\av}[1]{\left\langle #1 \right\rangle}	
\newcommand{\avb}[1]{\big\langle #1 \big\rangle}	
\newcommand{\eq}{\mathrm{eq}}
\newcommand{\s}[1]{\widetilde #1}
\newcommand{\DeltaM}{\Delta m_{\chi \sbot}}
\newcommand{\Ysbot}{Y_{\sbot}}
\newcommand{\mx}{\ensuremath{m_{\chi}}}
\newcommand{\heff}{\ensuremath{h_{\text{eff}}}}
\newcommand{\geff}{\ensuremath{g_{\text{eff}}}}

\newcommand{\sbotb}{\ensuremath{\tilde{b}}}

\def\beq{\begin{equation}}
\def\eeq{\end{equation}}
\definecolor{darkgreen}{RGB}{0,170,0}
\definecolor{darkgray}{RGB}{110,110,108}

\newcolumntype{C}{>{$}c<{$}} 	

\begin{document}

\title{Coannihilation without chemical equilibrium}

\author{Mathias Garny}
\affiliation{Physik Department T31, Technische Universit\"at M\"unchen,
James-Franck-Stra\ss e 1,
D-85748 Garching, Germany}
\author{Jan Heisig}
\affiliation{Institute for Theoretical Particle Physics and Cosmology, RWTH Aachen University, Sommerfeldstra\ss e 16, D-52056 Aachen, Germany}
\author{Benedikt L\"ulf}
\affiliation{Institute for Theoretical Particle Physics and Cosmology, RWTH Aachen University, Sommerfeldstra\ss e 16, D-52056 Aachen, Germany}
\author{Stefan Vogl}
\affiliation{Max-Planck-Institut f\"ur Kernphysik,\\ Saupfercheckweg 1, D-69117 Heidelberg, Germany}
\preprint{TUM-HEP 1085/17}
\preprint{TTK-17-18}

\begin{abstract}
Chemical equilibrium is a commonly made assumption in the freeze-out calculation of coannihilating dark matter. We explore the possible failure of this assumption and find a new conversion-driven freeze-out mechanism. Considering a representative simplified model inspired  by supersymmetry with a neutralino- and sbottom-like particle we find regions in parameter space with very small couplings accommodating the measured relic density. In this region freeze-out takes place out of chemical equilibrium and dark matter self-annihilation is thoroughly inefficient. The relic density is governed primarily by the size of the conversion terms in the Boltzmann equations. Due to the small dark matter coupling the parameter region is immune to direct detection but predicts an interesting signature of disappearing tracks or displaced vertices at the LHC. Unlike freeze-in or superWIMP scenarios, conversion-driven freeze-out is not sensitive to the initial conditions at the end of reheating.
\end{abstract}

\maketitle

\section{Introduction}\label{sec:intro}

The origin and the nature of dark matter (DM) in the Universe is one of the most pressing questions in 
particle- and astrophysics. Despite impressive efforts to uncover its interactions with the Standard Model (SM) of particle physics in (in)direct detection and accelerator-based experiments, 
DM remains elusive and, so far, our understanding is essentially limited to its gravitational interactions (see e.g.~\cite{Bertone:2004pz,Arcadi:2017kky}). 
It is therefore of utmost interest to 
investigate mechanisms for the generation of DM in the early Universe that go beyond the widely studied paradigm of thermal freeze-out, and that can point towards non-standard signatures. 

In this spirit we subject the well-known coannihilation scenario~\cite{Griest:1990kh} to further scrutiny and investigate the importance of the commonly made assumption of chemical equilibrium (CE) between DM and the coannihilation partner. This requires solving the full set of coupled Boltzmann equations which has been performed in the context of specific supersymmetric scenarios~\cite{Chung:1997rq,Ellis:2015vaa}. Here we consider a simplified DM model and explore the break-down of
CE in detail finding a new, conversion-driven solution for DM freeze-out which points towards a small interaction strength of the DM particle with the SM bath.
While the smallness of the coupling renders most of the conventional signatures
of DM unobservable, new opportunities for collider searches arise. In particular we find that searches for long-lived particles at the LHC are very powerful tools for testing conversion-driven freeze-out.

The structure of the paper is as follows: We begin by introducing a simplified model for coannihilations in Sec.~\ref{sec:model}. In Sec.~\ref{sec:BMEs} we present the Boltzmann equations which govern the DM freeze-out and investigate conversion-driven solutions before we confront the regions of parameter space which allow for a successful generation of DM with LHC searches in Sec.~\ref{sec:pheno}. Finally, we summarize our results and conclude. 

In the appendix we provide further details and results. In particular, we discuss the appearance of divergencies in the conversion rates in Appendix~\ref{Appx:rates}, describe our treatment of Sommerfeld enhancement in Appendix~\ref{sec:sommer} and justify the assumption of kinetic equilibrium by providing solutions of the full momentum-dependent 
Boltzmann equation in Appendix~\ref{sec:kineq}.

\section{Simplified model for coannihilation}\label{sec:model}

While the precise impact of the breakdown of CE between DM and its coannihilation partner will in general depend  on the details of the considered model, the key aspects of the phenomenology can be expected to be universal.   
As a representative case we choose a simplified model for DM interacting with quarks. 
We extend the matter content of the SM minimally by a Majorana fermion $\chi$, being a singlet under the SM gauge group, and a scalar quark-partner  $\s{q}$, mediating the interactions with the SM and acting as the coannihilation partner.
The interactions of the new particles among themselves and with the SM are given by~\cite{Garny:2012eb} 
\begin{equation}
	\mathcal{L}_\text{int} = |D_\mu \s{q}|^2 - \lambda_{\chi} \s{q} \bar{q}\frac{1-\gamma_5}{2}\chi +\text{h.c.},
	\label{eq:sbottommodel}
\end{equation}
where $q$ is a SM quark field, $D_\mu$ denotes the covariant derivative, which contains the interactions of $\s{q}$ with the gauge bosons as determined by its quantum numbers, and $\lambda_{\chi}$ is a Yukawa coupling.  
Here we choose $q=b$ and $Y=-\frac{1}{3}$. For the coupling $\lambda_{\chi} =\frac{1}{3}\sqrt{2}\frac{e}{\cos \theta_W} \approx 0.17$ our simplified model makes contact with the Minimal Supersymmetric SM (MSSM) where $\sbot$ can be identified with a right-handed sbottom and $\chi$ with a bino-like neutralino. However, here we will not consider the MSSM but vary $\lambda_{\chi}$ freely.\footnote{For a realization of small $\lambda_\chi$ in extensions of the MSSM see~\cite{Ibarra:2008kn}.}
Note that choosing a top-partner mediator instead yields similar results although  quantitative differences arise due to the large top mass.

On top of the gauge and Yukawa interactions described above a Higgs-portal interaction given by 
\begin{align}
\mathcal{L}_{h}=\lambda_h h^\dagger h \sbot^\dagger \sbot\,
\end{align}
is also allowed. This interaction does not involve $\chi$ directly and has no impact on the conversion rates $\chi \leftrightarrow\sbot$, that are responsible for establishing CE. Nevertheless, it can  modify the annihilation rate of $\sbot$. Since the additional contributions involving the scalar coupling compete with QCD processes, \emph{i.e.} $\sbot \sbot^\dagger\rightarrow g g$, they are sub-leading unless $\lambda_h$ is very large. Even in this case we do not expect qualitative differences, and therefore neglect this contribution in the following.

\section{Freeze-out without chemical equilibrium}\label{sec:BMEs}

For coannihilation to be effective the coannihilating particles -- here $\chi$ and $\sbot$ -- 
have to be in thermal contact through efficient 
conversion rates $\chi \leftrightarrow\sbot$. For couplings $\lambda_\chi$ of the order of the electroweak coupling strength, conversion rates  are much larger than the Hubble rate $H$ during freeze-out, guaranteeing CE, \emph{i.e.}\begin{align}
\frac{n_{\chi}}{n_{\chi}^{\eq}}=\frac{n_{\sbot}}{n_{\sbot}^{\eq}}\,,
\end{align}
where $n$ ($n^{\eq}$) is the (equilibrium) number density.
While CE holds the results are not sensitive to the size of the conversion rates and the Boltzmann equations can be reduced to a single equation that does not contain conversion terms~\cite{Edsjo:1997bg}. This approach is solved in standard tools~\cite{Belanger:2004yn,Gondolo:2004sc,Backovic:2013dpa}.

For smaller couplings, however, CE can break down and the full coupled set of Boltzmann equations has to be solved~\cite{Chung:1997rq,Ellis:2015vaa}. In our case:
\begin{widetext}
\begin{eqnarray}
		\frac{\diff Y_{\chi }}{\diff x} = \frac{1}{ 3 H}\frac{\mbox{d} s}{\mbox{d} x}&\Bigg[&\avb{\sigma_{\chi\chi}v}\left(Y_{\chi}^2-Y_{\chi}^{\eq\,2}\right)+\avb{\sigma_{\chi\sbot} v}\left(Y_{\chi}Y_{\sbot{}}-Y_{\chi}^{\eq}Y_{\sbot{}}^{\eq}\right)  \nonumber \\
		 &&+\;\frac{\Gamma_{\chi\rightarrow \sbot{}}}{s}\left(Y_{\chi}-Y_{\sbot{}}\frac{Y_{\chi}^{\eq}}{Y_{\sbot{}}^{\eq}}\right) -\frac{\Gamma_{\sbot} }{s}\left(Y_{\sbot{}}-Y_{\chi}\frac{Y_{\sbot}^{\eq}}{Y_{\chi}^{\eq}}\right)+\avb{\sigma_{\chi\chi\rightarrow \sbot{}\sbot{}^\dagger}v}\left(Y_{\chi}^2-Y_{\sbot}^2\frac{Y_{\chi}^{\eq\,2}}{\Ysbot^{\eq\,2}}\right) \Bigg] \label{eq:BMEchi}\\
		\frac{ \diff Y_{\sbot } }{\diff x} = \frac{1}{ 3 H}\frac{\mbox{d} s}{\mbox{d} x}&\Bigg[& \frac{1}{2}\avb{\sigma_{\sbot{}\sbot{}^\dagger}v}\left(Y_{\sbot{}}^2-Y_{\sbot{}}^{\eq\,2}\right)+\avb{\sigma_{\chi \sbot{}}v}\left(Y_{\chi}Y_{\sbot{}}-Y_{\chi}^{\eq}Y_{\sbot{}}^{\eq}\right) \nonumber\\ 
		 &&-\;\frac{\Gamma_{\chi\rightarrow \sbot{}}}{s}\left(Y_{\chi}-Y_{\sbot{}}\frac{Y_{\chi}^{\eq}}{Y_{\sbot{}}^{\eq}}\right) +\frac{\Gamma_{\sbot{}}}{s}\left(Y_{\sbot{}}-Y_{\chi}\frac{Y_{\sbot{}}^{\eq}}{Y_{\chi}^{\eq}}\right)
		 -\avb{\sigma_{\chi\chi\rightarrow \sbot{}\sbot{}^\dagger}v}\left(Y_{\chi}^2-Y_{\sbot{}}^2\frac{Y_{\chi}^{\eq\,2}}{Y_{\sbot{}}^{\eq\,2}}\right) \Bigg]\label{eq:BMEsbot}\;,
\end{eqnarray}
\end{widetext} 
where $Y= n/s$ is the comoving number density, $s$ the entropy density and $x=m_\chi/T$.
We take the internal degrees of freedom $g_{\chi}=2$ and $g_{\sbot}=3$. 
 $Y_{\sbot}$ represents the summed contribution of the mediator and its anti-particle. Since the cross sections are averaged over initial state degrees of freedom, this leads to the factor $1/2$ in the respective Boltzmann equation,
 Eq.~\eqref{eq:BMEsbot}. Equally, 
$\Gamma_{\chi\rightarrow \sbot}$ is understood to contain the conversion into both.

Apart from the familiar annihilation and coannihilation terms displayed in the first lines of Eqs.~\eqref{eq:BMEchi} and \eqref{eq:BMEsbot} three additional rates for the conversion processes enter in the second lines. The first term includes all the scattering processes which convert DM in its coannihilation partner. The  scattering rate is given by
\beq
\Gamma_{\chi\rightarrow \sbot}=2 \sum_{k,l}\big \langle\sigma_{\chi k  \rightarrow \sbot l}v \big\rangle \,n_k^{\eq}\,,
\eeq
where $k,l$ denote SM particles.
The factor of two arises from annihilation into the mediator and its anti-particle.
Neglecting quantum statistical factors and assuming Boltzmann distributions the thermal average for scattering reads~\cite{Edsjo:1997bg}
\beq
\begin{split}
&\av{\sigma_{ij} v}n_i^{\eq}n_j^{\eq}=  \\
&\qquad\; T \frac{g_i g_j}{256\pi^5}\int \frac{p_{ij}p_{ab}}{\sqrt{s}}|\overline{M}|^2K_1\!\left(\frac{\sqrt{s}}{T}\right) \diff s \,\diff \cos \theta\;\,,
\end{split}
\eeq
where $K_i$ denotes a  modified Bessel function of the second kind.
Here $g_i$ are the internal degrees of freedom of species $i$, $p_{ij}$ and $p_{ab}$ denote the absolute value of the three momentum of the initial and final state particles in the center-of-mass frame, respectively. In certain scattering processes soft- or $t$-channel divergences can appear in the thermal averages. We discuss these issues and how we resolve them in detail in Appendix~\ref{Appx:rates}.
The next term in the second lines of Eqs.~\eqref{eq:BMEchi} and \eqref{eq:BMEsbot} captures the conversion induced by the decay and inverse decay of $\sbot$. This rate is controlled by the thermally averaged decay width $\Gamma_{\sbot}$. The thermally averaged decay rate is given by 
\beq
	\Gamma_{\sbot} \equiv \Gamma \av{\frac{1}{\gamma}} = \Gamma \frac{K_1\left( \msbot /T\right) }{ K_2 \left( \msbot /T \right) }\;,
\eeq
where $\gamma$ is the Lorentz factor.
Finally, the last term takes the scattering processes in the odd-sector into account. We include all diagrams that are allowed at tree-level and
use \textsc{FeynRules}~\cite{Alloul:2013bka} and \textsc{CalcHEP}~\cite{Belyaev:2012qa} to  generate the squared matrix elements $|\overline{M}|^2$, see Tables~\ref{tab:coannihillations} and \ref{tab:conversions} for a summary of the included processes. We take into account the Sommerfeld enhancement of the $\sbot \sbot^\dagger$ annihilation rates
as detailed in Appendix~\ref{sec:sommer}.

\renewcommand{\arraystretch}{1.5} 

\begin{table}
	\centering
	\begin{tabular}{|C|C|C|C|C|}\hline
	\multicolumn{2}{|c|}{initial state}&\multicolumn{2}{c|}{final state}&\text{scaling}\\ \hline\hline
	\chi & \chi & b & \bar{b} &\lambda_\chi^4 \\ \hline\hline
	\multirow{2}{1cm}{\centering $\chi$} & \multirow{2}{1cm}{\centering $\sbot$} &\multirow{1}{1cm}{\centering $b$}&g,\gamma,Z,H &\multirow{2}{*}{$\lambda_\chi^2$} \\ \cline{3-4}
	 &  &W^-&t,u,c & \\ \hline\hline
	\multirow{4}{*}{$\sbot$} & \multirow{4}{*}{$\sbot^\dagger$} &V&V&\multirow{4}{*}{$\lambda_\chi^0$}\\ \cline{3-4}
	 &  &q&\bar{q} & \\ \cline{3-4}
	 &  &Z&H & \\ \cline{3-4}
	 &  &{l}&{\bar{l}} & \\ \hline
	\sbot & \sbot &b&b & \lambda_\chi^4 \\ \hline
	\end{tabular}
	\caption{List of all included coannihilation processes. We use the abbreviations $q=$ all quarks, $l=$ all leptons 	and $V=g,\gamma,Z,W$. The $\sbot$ annihilation into $b\bar{b}$ also has contributions scaling with $\lambda_\chi^2$ and $\lambda_\chi^4$.
	}
	\label{tab:coannihillations}
\end{table}

\begin{table}
	\begin{tabular}{|C|C|C|C|C|C|} \hline
		\multicolumn{2}{|c|}{initial state} & \multicolumn{2}{c|}{final state} & \text{symbol}&\text{scaling}  \\ \hline\hline
		\multirow{4}{0.7cm}{\centering $\chi$}& b &\multirow{4}{0.7cm}{\centering $\sbot$} & {g^*,\gamma^*,Z,H}&\multirow{4}{*}{$\Gamma_{\chi\rightarrow \sbot}$}&\multirow{4}{*}{$\lambda_\chi^2$}\\ \cline{2-2}\cline{4-4}
		& g,\gamma,Z^{**},H^{**} & & \bar{b} &&\\\cline{2-2}\cline{4-4}
		& W^- & & \bar{t},{\bar{u}^{**},\bar{c}^{**}} &&\\\cline{2-2}\cline{4-4}
		& {t^{**}},u,c & & W^+ && \\\hline\hline
		\multicolumn{2}{|C|}{\sbot} &\chi &b& \Gamma_{\sbot}&\lambda_\chi^2 \\\hline\hline
		\chi& \chi &\sbot & \sbot^\dagger& \langle \sigma_{\chi\chi \rightarrow \sbot\sbot^{\dagger}}v \rangle &\lambda_\chi^4\\\hline
	\end{tabular}
	\caption{List of all considered conversion processes. 
	Processes marked with ${}^*$ have soft divergences, processes with ${}^{**}$ can have $t$-channel divergences.}
	\label{tab:conversions}
\end{table}

A first naive estimate that allows us to determine the ballpark of the parameters where conversion processes become relevant follows from demanding
\beq
\Gamma_{\sbot} \frac{Y_{\sbot}^{\eq}}{Y_\chi^{\eq}} \sim H\;,
\label{eq:rateestimate}
\eeq
for temperatures relevant to the freeze-out dynamics ($m_\chi/T\sim 30$).
Using  as a representative benchmark $m_\chi=500\,$GeV and a value of the mediator mass that allows for coannihilations ($m_{\sbot}= 510$\,GeV) this relation indicates $\lambda_\chi \sim \mathcal{O}(10^{-7})$.  
The order of magnitude is largely insensitive to the precise choice of masses, as long as coannihilations can occur.

\begin{figure*}[t]
\centering
\setlength{\unitlength}{1\textwidth}
\begin{picture}(0.96,0.31)
  \put(0.0,-0.008){\includegraphics[width=0.96\textwidth]{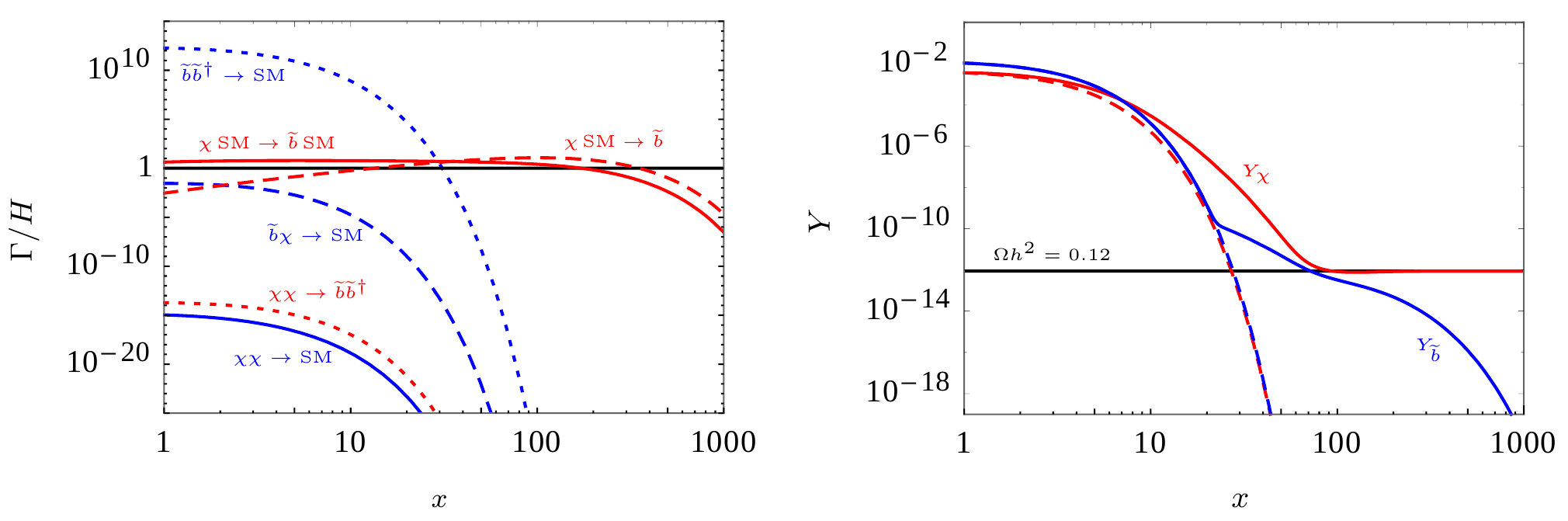}}
\end{picture}
\caption{Left panel: Rates of annihilation (blue curves) and conversion (red curves) terms in the Boltzmann equation relative to
the Hubble rate as a function of
$x=m_\chi/T$ for $m_\chi=500\,$GeV, $\msbot=510\,$GeV, $\lambda_\chi\approx 2.6\times 10^{-7}$. 
Right~panel:~Evolution of the resulting abundance (solid curves) of $\s b$ (blue) and $\chi$ (red). The dashed curves denote the equilibrium abundances.
}
\label{fig:ratesY}
\end{figure*}

For such a small coupling  a clear hierarchy emerges between the different rates, see left panel of Fig.~\ref{fig:ratesY}. The annihilation $\chi \chi \rightarrow$ SM SM and $\chi \chi \rightarrow \sbot \sbot^\dagger$ that are proportional to $\lambda_\chi^4$ and thermally suppressed by $n_{\chi}^{\eq}$ are exceedingly small and cannot compete with the Hubble expansion. 
Even though the coannihilation rate $\chi \sbot \rightarrow $ SM SM, which scales as  $\sigma_{\chi \sbot} v \propto \lambda_\chi^2 g^2$ (where $g$ is a SM gauge coupling) is enhanced relative to this by many orders of magnitude
 it is also negligible compared to $H$. In contrast, the leading contribution to $\sbot \sbot^\dagger \rightarrow$ SM SM is set by the gauge interactions of $\sbot$ and, therefore, the rate remains comfortably larger than $H$ until 
$T\approx m_{\chi}/30$. 
The most important annihilations, especially for very small $\lambda_\chi$ are the $\sbot$ annihilations into gluons. 
Since the interaction rates are suppressed exponentially by the masses of external particles, it is clear that the conversion processes containing external gluons dominate over the rates containing weak scale particles. 
 The conversion rates  are close to the Hubble rate and, for this choice of couplings, just about sufficient to make conversion processes relevant for the freeze-out.

Taking these rates and solving the Boltzmann equations we find the results presented in the right-hand side of Fig.~\ref{fig:ratesY}.
We solve the system of coupled equations from $x=1$ up to $x=1000$.\footnote{Due to efficient annihilations, the $\sbot$ abundance is
very close to equilibrium at early times. For numerical convenience, it is sufficient to track its deviation 
from equilibrium starting from $x\sim 15$.}
The $\chi$ abundance leaves its equilibrium value already at rather high temperatures, well before the freeze-out of a typical thermal relic or the $\sbot$ freeze-out. The slow decline of the $\chi$ abundance after this point is  due to the close-to-inefficient conversion terms which remove over-abundant $\chi$s.

\begin{figure}[b]
\centering
\setlength{\unitlength}{1\textwidth}
\begin{picture}(0.47,0.32)
  \put(0.0,-0.008){\includegraphics[width=0.46\textwidth]{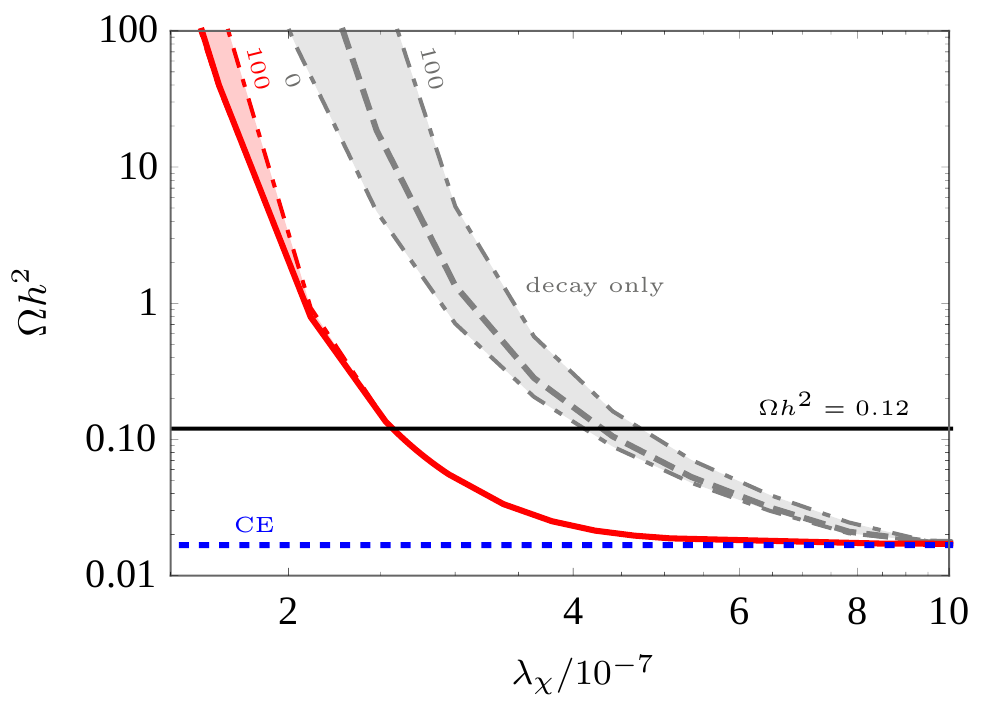}}
\end{picture}
\caption{Relic density as a function of the coupling $\lambda_\chi$, for $m_\chi=500\,$GeV, $\msbot=510\,$GeV. The dotted blue line is the result that would be obtained when assuming CE.
 The red line shows the full solution including all conversion rates, the gray dashed line corresponds to the solution when only decays are considered. The shaded areas highlight the dependence on initial conditions,
$Y_{\chi}(1)=(0\!-\!100)\times Y_{\chi}^{\eq}(1)$.
The central curves correspond to $Y_{\chi}(1)=Y_{\chi}^{\eq}(1)$.
}
\label{fig:coupling}
\end{figure}

In Fig.~\ref{fig:coupling} we show the dependence of the final freeze-out density on the coupling $\lambda_\chi$ (red solid line).
For large enough coupling, the solution coincides with the result that would be obtained when assuming CE (blue dotted
line). The relic density is in this case largely set by the strength of $\sbot$ self-annihilation into gluons. When lowering the value of $\lambda_\chi$, conversions $\chi\leftrightarrow\sbot$ become less efficient and one obtains a relic density that lies above the value expected for CE\@. 
For the benchmark scenario shown in Fig.~\ref{fig:coupling}, the freeze-out density matches the value determined by Planck~\cite{Ade:2015xua}
for a coupling of $\lambda_\chi\approx 2.6 \times 10^{-7}$.

\begin{figure}[b]
\centering
\setlength{\unitlength}{1\textwidth}
\begin{picture}(0.46,0.315)
  \put(0.00,-0.008){\includegraphics[width=0.46\textwidth]{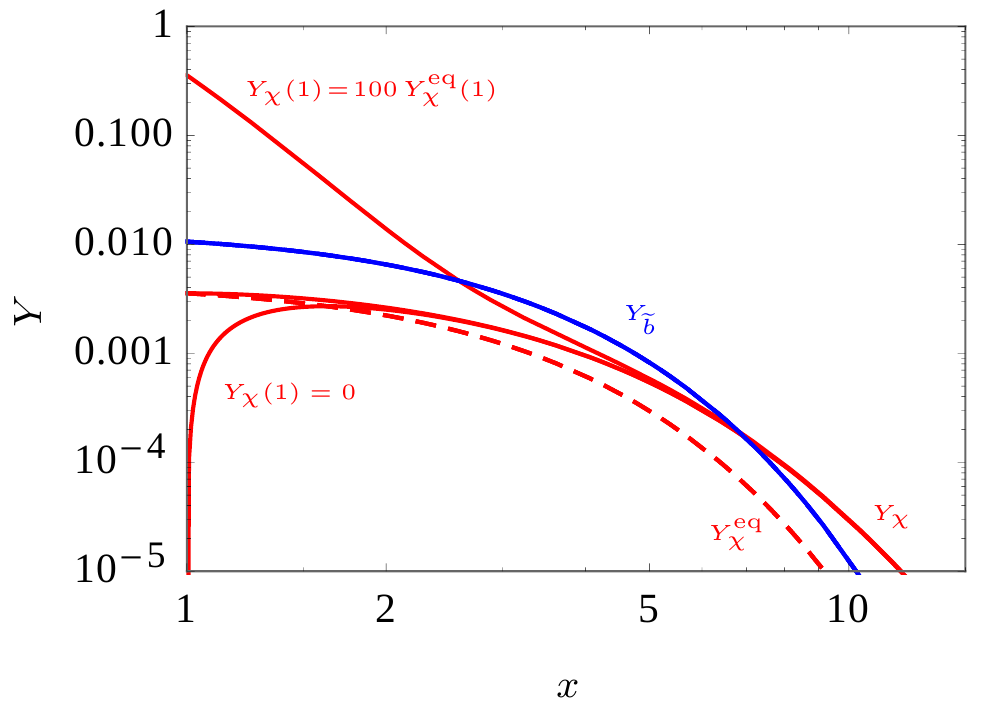}}
\end{picture}
\caption{Dependence on the initial conditions for $Y_{\chi}$ at $x=1$. We show solutions for 
the choices $Y_{\chi}(1)=[0,1,100]\times Y_{\chi}^{\eq}(1)$,
and otherwise the same parameters as in Fig.~\ref{fig:ratesY}.}
\label{fig:initialdep}
\end{figure}

Above we assumed that both $\chi$ and $\sbot$ have thermal abundances for $T\gg m_\chi$. While this assumption is certainly well justified for $\sbot$,
one may question the dependence on the initial condition for $\chi$ due to its small coupling to the thermal bath.
We check the dependence on this assumption by varying the initial abundance at $T=m_\chi$ within the range $(0\!-\!100)\times Y_\chi^{\eq}$. The evolution of the abundances for our benchmark point are shown in Fig.~\ref{fig:initialdep}, for early times ($x<20$). We find that all trajectories converge before $x\lesssim 5$, thereby effectively
removing any dependence of the final density on the initial condition at $x=1$.
The dependence of the final freeze-out density on the initial condition is also indicated  in Fig.~\ref{fig:coupling} by the area shaded in red, and is remarkably small.
Therefore, conversion-driven freeze-out is largely insensitive to details of the thermal history  prior to freeze-out and in particular to a potential production during the reheating process. Note that this feature distinguishes conversion-driven freeze-out from scenarios for which DM has an even weaker coupling such that it was never in thermal contact (e.g. freeze-in production~\cite{Hall:2009bx}). Thus, while requiring a rather weak coupling, the robustness of the conventional freeze-out paradigm is preserved in the scenario considered here.

As discussed before, conversions $\chi \leftrightarrow \sbot$ are driven by two types of processes, decay and scattering. It turns out that
quantitatively both are important for determining the freeze-out density. To illustrate the importance of scattering processes, we show the
freeze-out density that would be obtained when only taking decays into account by the gray dashed line in Fig.~\ref{fig:coupling}.
Furthermore, the gray shaded area indicates the dependence on initial conditions that would result neglecting scatterings.
We find that scattering processes, which are active at small $x$, are responsible for wiping out the dependence on the initial abundance
in the full solution of the coupled Boltzmann equations.

While Eqs.~\eqref{eq:BMEchi} and \eqref{eq:BMEsbot} do not require chemical equilibrium to hold, they rely on the implicit assumption of kinetic
equilibrium, \emph{i.e.} the assumption that the momentum dependence of the distribution functions is proportional to a thermal distribution. For the freeze-out of weakly interacting massive particles, this assumption
is typically well justified because scattering processes are enhanced compared to annihilations by a Boltzmann factor of order $e^{m_\chi/T}$.
However, in the present scenario, the small coupling $\lambda_\chi$ renders elastic scatterings of the form $\chi b\leftrightarrow \chi b$
inefficient. Instead, the leading processes to establish kinetic equilibrium are the inelastic conversion processes discussed above.
Since their rate is, by definition, comparable to the Hubble rate in the interesting regime of parameters, one may wonder whether the
treatment based on integrated Boltzmann equations is justified for $\chi$. In order to check this point, we solved the full, momentum-dependent
Boltzmann equation for $\chi$, taking the leading decay and scattering processes into account. We find that, while the distribution function can indeed
deviate from the thermal distribution at intermediate times, the final relic abundance differs only mildly from the integrated treatment (below the 10\%
level). The main
reason is that the collision operator does not depend strongly on the momentum mode, such that all modes behave in a similar way. 
For a detailed discussion we refer to appendix~\ref{sec:kineq}.

Let us briefly comment on possible refinements. Apart from quantum statistics, also thermal effects could play a role
at small $x$. In particular, the thermal mass for the $b$-quark can lead to a thermal blocking of the decay at high temperatures
and for very small mass splitting. Since a consistent inclusion of this effect would require to take also further thermal processes
into account, and since (hard) scatterings dominate for small $x$, we do not expect these corrections to significantly affect our
conclusion. Additionally, bound state effects could play a role for the $\sbot$ annihilation~\cite{Kim:2016zyy, Kim:2016kxt,Liew:2016hqo,Mitridate:2017izz,Keung:2017kot}.

\section{Viable parameter space}\label{sec:pheno}

\begin{figure}[b]
\centering
\setlength{\unitlength}{1\textwidth}
\begin{picture}(0.47,0.45)
  \put(0.00,-0.008){\includegraphics[width=0.47\textwidth]{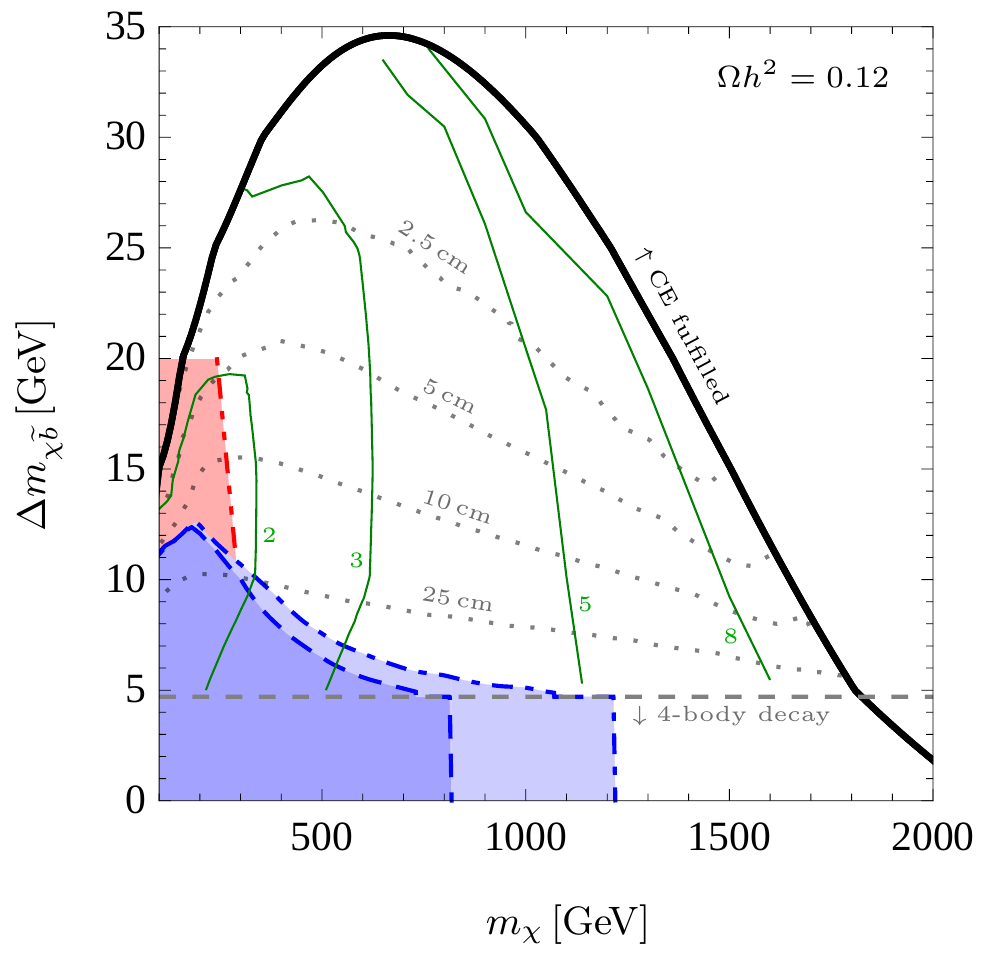}}
\end{picture}
\caption{
Viable parameter space in the plane spanned by $m_\chi$ and $\DeltaM=\msbot-m_\chi$. 
We adjust $\lambda_\chi$ such that $\Omega h^2=0.12$. 
Above the thick black curve CE holds, while
below this curve CE breaks down and the freeze-out is conversion-driven.
The corresponding coupling $\lambda_\chi/10^{-7}$ (decay length $c\tau$) of the mediator is denoted
by the thin green (gray) dotted lines.
The blue dashed (dot-dashed) curve shows our estimates for the limits from
$R$-hadron searches at 8 (13)\,TeV. The constraint from monojet searches is shown as the red dot-dot-dashed curve.  
}
\label{fig:regplot}
\end{figure}

We will now explore the parameter space consistent with a relic density that matches the DM density measured by Planck, $\Omega h^2 = 0.1198\pm 0.0015$~\cite{Ade:2015xua}.
In the considered scenario, for small couplings, $\sbot\sbot^\dagger$ annihilation is the only efficient annihilation channel. Hence the minimal relic density that can be obtained for a certain point in the $m_\chi$-$m_{\sbot}$ plane is the one for a coupling $\lambda_{\chi}$ that just provides CE (but is still small enough so that $\chi\chi$- and $\chi\sbot$-annihilation is negligible). The curve for which this choice provides the right relic density defines the boundary
of the valid parameter space and is shown as a black, solid curve in Fig.~\ref{fig:regplot}.
Below this curve a choice of $\lambda_{\chi}$ sufficiently large to support CE 
would undershoot the relic density. In this region a solution with small $\lambda_{\chi}$ exists
that renders the involved conversion rates just large enough to allow for the right 
portion of thermal contact between $\sbot$ and $\chi$ to provide the right relic density.
The value of $\lambda_{\chi}$ ranges from $10^{-7}$ to $10^{-6}$ (from small to large $m_\chi$).
These values lie far beyond the sensitivity of direct or indirect detection experiments.

For the solutions providing the right relic density, during typical freeze-out (\emph{i.e.}~when $T\sim m_\chi/30$) the conversion rates have to be on the edge of being efficient, \emph{cf.}~Eq.~\eqref{eq:rateestimate}. From this simple relation (and assuming that the decay width, $\Gamma_{\sbot}$, is similar in size as the other conversion rates) we can already infer that the decay length of $\sbot$ is of the order of 1--100\,cm for a DM particle with a mass of a few hundred GeV predicting the signature of disappearing tracks or displaced vertices at the LHC\@.

The decay length in our model is shown as the gray dotted lines in Fig.~\ref{fig:regplot}. It ranges from $25\,\text{cm}$ to below $2.5\,\text{cm}$ for increasing mass difference (the dependence on the absolute mass scale is more moderate).

In proton collisions at the LHC pairs of $\sbot$s could be copiously produced. They will 
hadronize to form $R$-hadrons~\cite{Farrar:1978xj} which will, for the relevant decay length, either decay inside
or traverse the sensitive parts of the detector. 
Accordingly, the signatures of displaced vertices and (disappearing) highly ionizing tracks provide promising discovery channels. 
Similar searches have, \emph{e.g.}, been
performed for a gluino $R$-hadron (decaying into energetic jets)~\cite{Aad:2015rba} or a purely electrically charged 
heavy stable particle~\cite{Aad:2013yna,CMS:2014gxa} but have not been performed
for the model under consideration (see also~\cite{Mahbubani:2017gjh,Buchmueller:2017uqu}
for a recent account on simplified DM models providing displaced vertices).
However, some constraints on the model can already be derived from existing searches.

Searches for detector-stable $R$-hadrons~\cite{Chatrchyan:2013oca,CMS-PAS-EXO-16-036,ATLAS:2014fka,Aaboud:2016dgf} can be reinterpreted for finite decay lengths by
convoluting the signal efficiency with the fraction of $R$-hadrons 
that decay after traversing the relevant parts of the detector. This reinterpretation provides limits
down to a decay length of $c\tau\simeq0.1\,$m for an $R$-hadron mass around 100\,GeV
and can be used to estimate excluded parameter regions in our model.

To this end we compute the weighted fraction of $R$-hadrons 
that decay after traversing the relevant parts of the detector in a Monte Carlo simulation as follows. 
For a given $R$-hadron in an event $i$ this fraction is 
\beq
\mathcal{F}^i_\text{pass} = e^{-\ell/\left(c\tau \beta\gamma\right)}\,,
\eeq
where $\ell=\ell(\eta)$ is the travel distance to pass the respective part of the detector
which depends on the pseudo-rapidity $\eta$ while $\gamma$
is the Lorentz factor according to the velocity $\beta$. We use a simple cylindrical approximation for the CMS 
tracker\footnote{We considered the tracker-only 
and tracker+muon-system analysis of~\cite{Chatrchyan:2013oca} finding the higher sensitivity
for the former one.} with a radius and length of 1.1\,m and 5.6\,m, respectively.
For the weighting we compute\footnote{%
For simplicity we display the formula for one $R$-hadron candidate per event, for events with two candidates 
we follow the prescription in~\cite{Khachatryan:2015lla} (with the replacement
$\mathcal{P}^i_\text{off} \to \mathcal{F}^i_\text{pass}\mathcal{P}^i_\text{off} $ in the respective 
sum in the numerator of Eq.~\eqref{eq:avFlong}).}
\beq
\label{eq:avFlong}
\overline{\mathcal{F}}_\text{pass}=
\frac{\sum_{i} \mathcal{F}^i_\text{pass} \mathcal{P}^i_\text{on} \mathcal{P}^i_\text{off}}
{\sum_{i} \mathcal{P}^i_\text{on} \mathcal{P}^i_\text{off}}\,,
\eeq
where $\mathcal{P}^i_\text{on}$ and $\mathcal{P}^i_\text{off}$ are
the probabilities of the respective event to be triggered and pass the selection cuts, respectively, and the sum
runs over all generated events. We use the tabulated probabilities \
$\mathcal{P}^i_\text{on}, \mathcal{P}^i_\text{off}$
for lepton-like heavy stable charged particles following the prescription
in~\cite{Khachatryan:2015lla} (see also~\cite{Heisig:2015yla} for details of the implementation of isolation criteria and validation). 
We expect this to be a good approximation as the selection criteria for lepton-like heavy stable charged particles and $R$-hadrons 
are identical and differences in the overall detector efficiency cancel out in Eq.~\eqref{eq:avFlong}.
We simulate events with \textsc{MadGraph5\_aMC@NLO}~\cite{Alwall:2014hca},
performing showering and hadronization with \textsc{Pythia~6}~\cite{Sjostrand:2006za}.

\begin{figure}[b]
\centering
\setlength{\unitlength}{1\textwidth}
\begin{picture}(0.47,0.44)
\put(0.0,-0.01){\includegraphics[width=0.46\textwidth]{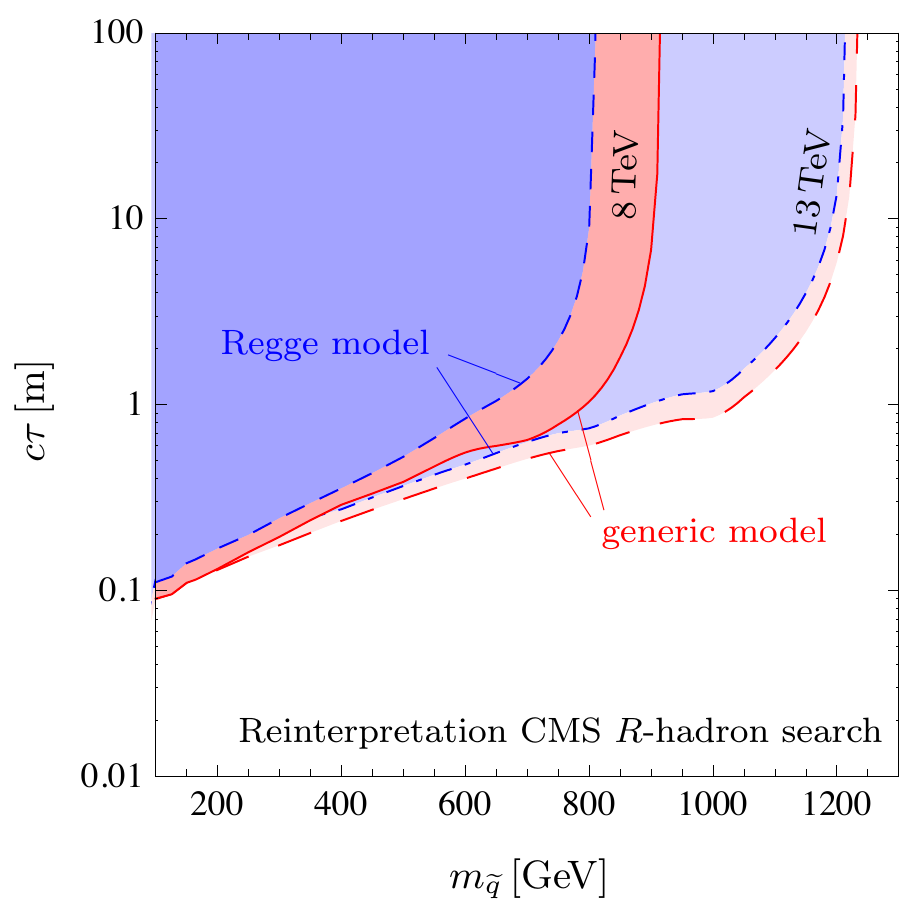}}
\end{picture}
\caption{Regions excluded at 95\% CL by a reinterpretation of the searches for detector stable 
top-squark $R$-hadrons with CMS at the 8\,TeV and 13\,TeV LHC 
(tracker-only analysis).}
\label{fig:LHClim}
\end{figure}

We use the cross section predictions from \textsc{NLLFast}~\cite{Beenakker:2010nq} and
rescale the signal by $\overline{\mathcal{F}}_\text{pass}$. The 95\% CL exclusion 
limits are then obtained from a comparison to the respective cross section limits 
from searches for (top-squark) $R$-hadrons presented in~\cite{Chatrchyan:2013oca}. 
The results are shown in Fig.~\ref{fig:LHClim}.
We show limits for two models regarding the hadronization and interaction of the $R$-hadron with the detector material, the generic model~\cite{Kraan:2004tz,Mackeprang:2006gx} and Regge (charge-suppressed) model~\cite{deBoer:2007ii,Mackeprang:2009ad} as the
red solid and blue dashed line, respectively.

In addition to the results for the 8\,TeV LHC we show results from an analogous reinterpretation
of the preliminary results from $12.9\,\text{fb}^{-1}$ of data from the 13\,TeV LHC 
run~\cite{CMS-PAS-EXO-16-036}.
Since the tabulated probabilities in~\cite{Khachatryan:2015lla} are only provided for 8\,TeV
we use these also for the analysis of the 13\,TeV simulation assuming a similar detector efficiency 
for $R$-hadrons in both runs.

The fraction of $R$-hadrons passing the tracker is exponentially suppressed for small 
life-times significantly weakening the respective sensitivity. However, there are two 
competing factors that nevertheless result in meaningful limits for $c\tau$ smaller than the detector
size. On the one hand, for small masses the production cross section rises quickly. On the other
hand, for smaller masses a larger fraction of $R$-hadrons is significantly boosted enhancing
the travel distance in the detector. However, this (latter) effect does not significantly enhance
the sensitivity as the signal efficiency for largely boosted $R$-hadrons 
decreases rapidly (as tracks become indistinguishable from minimal ionizing tracks for $\beta\to1$).

Note that the above CMS analysis has been interpreted for $R$-hadrons containing top-squarks. 
As discussed in~\cite{Mackeprang:2009ad} the expected energy loss for an $R$-hadron containing sbottoms
is smaller. This results in an efficiency around 30--40\% smaller relative to the case of the stop and therefore in 
slightly weaker limits on the sbottom mass, see \emph{e.g.}~\cite{ATLAS:2014fka}. 
However, we use the above limit taking the result for the Regge model
(that provides the weaker limits) as a realistic
estimate of the LHC limits for the (sbottom-like) mediator in our model
considering the fact that the uncertainties in the hadronization 
are of similar size as the difference between the sbottom and stop case.

The resulting limits from the 8\,TeV~\cite{Chatrchyan:2013oca}
and 13\,TeV~\cite{CMS-PAS-EXO-16-036} LHC data are superimposed in Fig.~\ref{fig:regplot}. For mass splittings below $m_b$ (below gray dashed curve)
the 2-body decay is not allowed and the $R$-hadrons can be considered detector-stable.
Towards large mass splittings (smaller life-times) the limits fall off significantly providing no 
constraint above $\DeltaM\simeq 13\,\text{GeV}$.

In addition, a large number of experimental results for a sbottom-neutralino
simplified model assuming a prompt sbottom decay exist, see~\emph{e.g.}~\cite{Sirunyan:2017cwe,CMS:2017kmd,Sirunyan:2016jpr,Aaboud:2016nwl}. While most of these searches are not applicable to non-prompt decays, monojet searches, targeting small mass splittings, have been performed 
that do not rely on the prompt decay of the mediator~\cite{Khachatryan:2015wza,Aaboud:2016tnv}.
We superimpose the (stronger) limit from~\cite{Aaboud:2016tnv} that uses $3.2\,\text{fb}^{-1}$ of 13\,TeV data.

\section{Conclusion}\label{sec:summary} 

In this work we have considered the possibility that the common assumption of chemical equilibrium during DM freeze-out does not hold.
For definiteness, we have focused on a simplified model with particle content inspired by supersymmetry, comprising
a neutral Majorana fermion as the DM candidate and a colored scalar particle that mediates a coupling to bottom quarks. For small mass splitting
between the mediator and the DM particle, the freeze-out is dominated by self-annihilation of the mediator. This process can be efficient enough
to deplete the DM density below the observed value, thus giving rise to a portion of parameter space in which thermal freeze-out cannot account
for all of the DM abundance. In this work we have demonstrated that this conclusion hinges on the assumption of chemical equilibrium, and that the freeze-out process can account for the DM density determined by Planck when relaxing this assumption.
This occurs when the DM particle interacts very weakly with both the SM and the mediator, such that conversion processes have to be taken into account
explicitly. 
We find that this opens up new regions in parameter space which lead to characteristic signatures of long-lived particles at collider experiments. $R$-hadron searches performed at the
$8$ and $13$\,TeV LHC runs already constrain part of the parameter space providing conversion-driven freeze-out. A dedicated search for disappearing $R$-hadron tracks and displaced vertices targeting decay lengths in the range 1--100 cm is expected to probe an even larger portion of the allowed parameter space.

The mechanism discussed here is distinct from the freeze-in scenario~\cite{Hall:2009bx}, for which the DM particle was never in thermal equilibrium,
and which would require a much smaller coupling strength than considered here. On the other hand, it shares some similarities with the
superWIMP scenario (see \emph{e.g.}~\cite{Feng:2003uy}), but also differs in various respects. In particular, the relic density is set by the interplay of conversion and
annihilation processes during freeze-out, unlike for superWIMPs, where DM is produced from the late decay
of a heavier state that
undergoes a standard thermal freeze-out. In addition, for the mechanism considered here, the life-time of the coannihilation partner is short enough
such that constraints from Big Bang nucleosynthesis are generally avoided, provided that the decay rate gives a sizable contribution to conversion.
Unlike both freeze-in and superWIMP scenarios in the considered mechanism the final relic density is insensitive to
the initial condition of the abundance at the end of the reheating process.

We expect that conversion-driven freeze-out can be realized generically in DM models featuring strong coannihilations.
If the coannihilation partner is not colored
but only electrically charged, one may expect signatures related to lepton-like highly ionizing tracks. Finally, it is possible that the efficient self-annihilation
of the coannihilation partner is itself driven by a new interaction beyond the SM~\cite{Baker:2015qna}. In this case the mechanism described here can be
relevant even if the coannihilating state is a SM singlet with macroscopic decay length, potentially leading to displaced vertex signatures.

\emph{Note added:} Reference~\cite{DAgnolo:2017dbv}, which appeared recently, discusses a related mechanism.

\section*{Acknowledgements}

We thank Michael Kr{\"a}mer, Bj\"orn Sarrazin, Pasquale Serpico and Wolfgang Waltenberger
for helpful discussions. 
We acknowledge support by the German Research Foundation DFG through the 
research unit ``New physics at the LHC''.

\begin{appendix}

\section{Divergences in conversion rates}\label{Appx:rates}

\begin{figure}[t]
\centering
\setlength{\unitlength}{1\textwidth}
\begin{picture}(0.46,0.32)
 \put(0.00,-0.01){\includegraphics[width=0.45\textwidth]{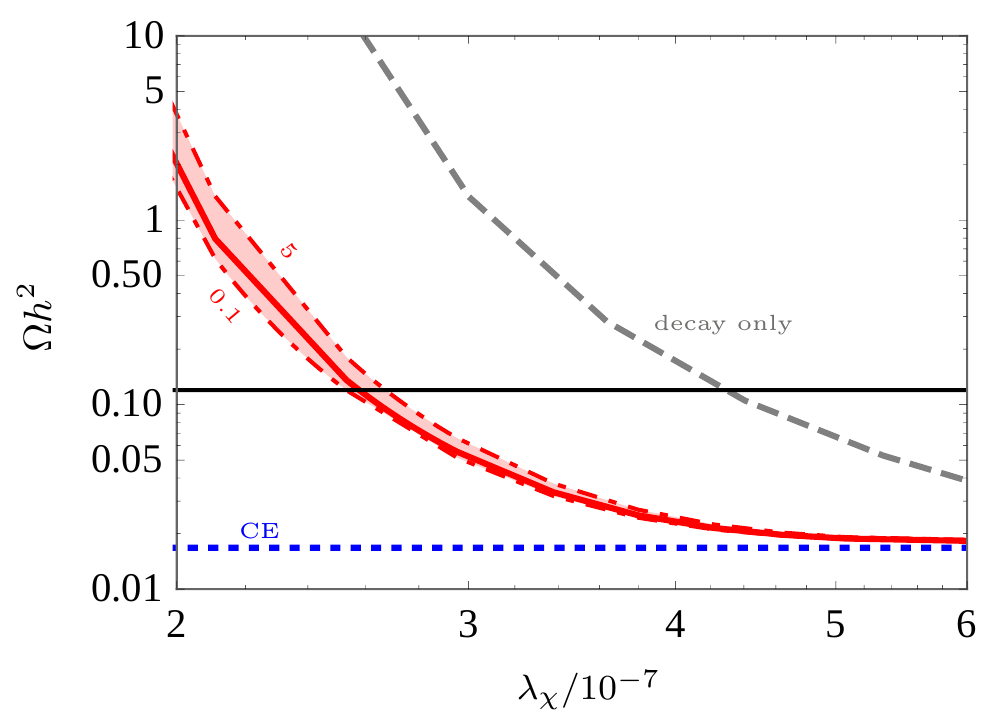}}
\end{picture}
\caption{Dependence of the final DM density on the regularization parameter 
$m_{\text{cut}}\in[0.1,0.5,5]\,\text{GeV}$, for $m_\chi=500\,$GeV, $\msbot=510\,$GeV. 
}
\label{fig:regularizationdep}
\end{figure}

Due to the inclusion of scattering processes, two issues arise in the thermal averages.
Since we do not consider loop corrections to the two-body decay or $1\to 3$ processes $\sbot\to \chi b g$, we cannot use them to cancel soft
contributions from $g \sbot \leftrightarrow b \chi$ scatterings (of course, the $\gamma$ scattering also has this problem, as noted by ${}^*$ in Table~\ref{tab:conversions}).
Instead, we regularize these processes by imposing a cut on the minimal process energy of $s_{\text{min}}=(\msbot+m_{\text{cut}})^2$, with fiducial value
 $m_{\text{cut}}=0.5\,\text{GeV}$. We checked that our numerical results are stable when varying $m_{\text{cut}}$ over a wide range, see Fig.~\ref{fig:regularizationdep},
indicating that the bulk of the scattering processes occurs at energies above the $b$ mass.
On top of this, we find that in processes of the type $\sbot \bar{b} \leftrightarrow \chi H$ the $b$-quark in the $t$-channel is allowed to go on-shell for some center-of-mass energies (the affected processes are marked with ${}^{**}$ in Tab.~\ref{tab:conversions}).
This corresponds to a double counting of the on-shell two-body decay. We choose to suppress the on-shell part by introducing a large
Breit-Wigner width for the $b$-quark, taking $\Gamma_b= m_b$ in our numerical calculations. 
Since this issue occurs only in processes involving weak scale particles ($H,W,Z$) and the scattering rate is dominated by gluons the  precise value for the width does not have an appreciable 
impact on the results. 

\section{Sommerfeld enhancement}\label{sec:sommer}

In the presence of light degrees of freedom  non-perturbative corrections to the annihilation rates are known to become relevant in the non-relativistic limit~\cite{Hisano:2003ec,Hisano:2004ds}. Between pairs of color charged particles the exchange of gluons generates a potential which modifies the wave function of the initial state particles and leads to a non-negligible correction of the tree-level cross section~\cite{deSimone:2014pda,Ibarra:2015nca,Hryczuk:2011tq,ElHedri:2016onc}.

To leading order the effect of the QCD potential can be described by a Coulomb-like potential~\cite{Fischler:1977yf}
\begin{align}
V(r)\approx \frac{\alpha_s}{2 r}  \left[ C_Q- C_R - C_{R'}  \right]
\end{align} 
where $C_R$ and $C_{R'}$ denote the Casimir coefficients of the incoming particles while $C_Q$ is the Casimir coefficient of the final state. 
For a general Coulomb-potential with $V(r)= \alpha/r$ the $s$-wave Sommerfeld correction factor $S_0$ is given by~\cite{deSimone:2014pda}
\begin{align}
S_0 =-\frac{\pi \alpha/\beta}{1-e^{\pi \alpha /\beta}}\;,
\label{eq:Somm}
\end{align}
where $\beta =v /2$ and the total annihilation cross section of particles moving in this potential is given by
 $\sigma_{Somm} = S_0 \cdot \sigma_{tree}$.\footnote{In principle the Sommerfeld factors have to be determined separately for each partial wave. For the model considered here the total Sommerfeld effect can be approximated to good accuracy by applying the $s$-wave correction to the full cross section.}
 For final states which are exclusively in a singlet, i.e. $ZZ,W^+ W^-,\gamma\gamma$, or an octet representation, i.e. $\gamma g, Z g$, the enhancement is given by Eq.~\eqref{eq:Somm} with $\alpha=-4/3\alpha_s $ or $ \alpha=1/6 \alpha_s$, respectively. The $gg$ final state is slightly more complicated since it can be in a singlet or octet representation. 
After summing over the different contributions the total Sommerfeld correction factor for this case reads~\cite{deSimone:2014pda}
\begin{equation}
	S_0\rightarrow \frac{2}{7}S_0\Bigg|_{\alpha=-4/3\alpha_s}+\frac{5}{7}S_0\Bigg|_{\alpha=1/6 \alpha_s}\;.
\end{equation} 
Since this channel dominates the annihilation rates by orders of magnitude, we only take  the correction for annihilation to gluons into account.

\section{Kinetic equilibrium}\label{sec:kineq}

In this section we compare solutions of the differential, momentum-dependent Boltzmann equation for $\chi$ with the
integrated Boltzmann equation used in the main text. The latter relies on the assumption of kinetic equilibrium.
This assumption may be questionable for the range of parameters we are interested in, because the leading interactions of $\chi$ are the conversion processes that become inefficient around the time of freeze-out.

If a particle $X$ is in kinetic equilibrium with the thermal bath of SM particles
at temperature $T$, its distribution function is given by
\begin{equation}
	f_{X}(p,t)=f_{X}^{\eq}(p,T)\frac{Y_{X}(t)}{Y_{X}^{\eq}(T)}\,,
	\label{eq:kineq}
\end{equation}
where $f_{X}(p,t)$ is the phase-space density.
We assume that due to efficient coupling to the SM, the mediator $\sbot$ is in kinetic equilibrium for all relevant times. When dropping the assumption of kinetic equilibrium for $\chi$, the unintegrated Boltzmann equation for $f_{\chi}$,
\begin{equation}
	\left( \partial_t -Hp\partial_p\right)f_{\chi}(p,t)=\frac{1}{E_{\chi}}C\left[f_{\chi}\right]\,,
	\label{eq:uiBE}
\end{equation}
has to be solved, see \emph{e.g.}~\cite{DAgnolo:2017dbv,Binder:2017rgn}.
The collision operator $C$ determines the differential interaction rate of 
each momentum mode of $\chi$. Introducing the notation 
\begin{equation}
	x(t,p)=\mx/T\,,\qquad q(t,p)=p/T\,,
\end{equation}
the Liouville operator on the left-hand side of Eq.~\eqref{eq:uiBE} 
can be brought into the form
\begin{equation}
\begin{split}
	\left( \partial_t -Hp\,\partial_p\right)
	&=\frac{H}{1-\frac{x}{3\heff}\frac{\diff \heff}{\diff x}} \left(x\partial_x+ \frac{x}{3\heff}\frac{\diff \heff}{\diff x}q\partial_q\right)\\
	&\approx H x\,\partial_x\,.
\end{split}
\end{equation}
In the last line we assume $\heff$ to be constant during freeze-out which is approximately 
satisfied above temperatures of $\mathcal{O}(\text{GeV})$. 
Here we adapt the notation of~\cite{Edsjo:1997bg} for the effective degrees of freedom
$\geff$ and $\heff$ associated with the energy and entropy densities, respectively.

\subsection{Collisional operator for conversion}

For the purpose of comparing the integrated with the differential Boltzmann equation,
 we focus on the dominant interaction processes that drive the distribution function towards
kinetic equilibrium. 
For the parameter range we are interested in, these are conversion processes, 
in particular (inverse) decays $\chi b \leftrightarrow \sbot $,
and, for earlier times, also inelastic scatterings 
$\chi A \leftrightarrow \sbot B$.
We neglect the small contributions from $\chi\chi$ annihilation, $\chi\sbot$ coannihilation and 
elastic scattering $\chi b \leftrightarrow \chi b$. 
Note that if we would take these additional processes into account, one expects
them to drive the distribution function closer to 
its equilibrium shape.
Therefore the following analysis may be regarded as a conservative estimate of the deviations from
kinetic equilibrium.
 
With these assumptions Eq.~\eqref{eq:uiBE} becomes a linear differential equation
and the remaining contributions to $C\left[f_{\chi}\right]$ are the (inverse) decay $\chi b \leftrightarrow \sbot $,
\begin{equation}
	C_{12}[f_{\chi}]= \frac{1}{2}\!\int\diff \Pi_{b} \diff \Pi_{\sbot} (2\pi)^4\delta^4\!\left(\sum p_i\!\right) |\overline M|^2\! \left[ f_{\sbot} - f_{\chi} f_b \right] ,
\end{equation}
and the inelastic scatterings 
$\chi A \leftrightarrow \sbot B$,
\begin{equation}
\begin{split}
	C_{22}[f_{\chi}]=& \frac{1}{2}\int\diff \Pi_{A} \,\diff \Pi_{\sbot}\, \diff \Pi_{B} \,(2\pi)^4\delta^4\left(\sum p_i\right) |\overline M|^2 \\ &\qquad\times \left[ f_{\sbot} f_B -f_{\chi} f_A \right]\,,
\end{split}
\end{equation}
where $\diff\Pi_X=g_X \diff^3p_X/\left((2\pi)^3E_X\right)$.
Apart from the different integrations, the terms in brackets differ from those in the case of the integrated Boltzmann equation. 
For the decay term we find
\begin{equation}
	f_{\sbot} - f_{\chi} f_b = f_{\sbot}^{\eq} \frac{Y_{\sbot}}{Y_{\sbot}^{\eq}} - f_{\chi} f_b^\eq=f_b^{\eq}\left( f_{\chi}^{\eq} \frac{Y_{\sbot}}{Y_{\sbot}^{\eq}}-f_{\chi} \right).
\end{equation}
using the relation of detailed balance for the equilibrium distributions.
A similar simplification can be performed for inelastic scattering
such that $C$ factorizes into
\begin{equation}
	C[f_{\chi}]= \widetilde{C}(q,x) \left(f_{\chi}^{\eq} \frac{Y_{\sbotb}}{Y_{\sbotb}^{\eq}}-f_{\chi}\right) E_\chi.
\end{equation}
As before, we neglect quantum statistical factors in the calculation.
The unintegrated Boltzmann equation \eqref{eq:uiBE} can hence be written in the form
\begin{equation}
	H x\partial_x f_{\chi}(q,x)= \widetilde{C}(q,x)\left(f_{\chi}^{\eq} \frac{Y_{\sbotb}}{Y_{\sbotb}^{\eq}}-f_{\chi}\right).
\end{equation}
This ordinary differential equation together with the boundary condition 
\begin{equation}
	f_{\chi}(q,x_0)=f^{\eq}_{\chi}(q,x_0)=\exp\left( -\sqrt{q^2+x_0^2} \right)
\end{equation}
can be solved with separation of variables and variation of constants. The result reads (\emph{cf.}~\cite{DAgnolo:2017dbv})
\begin{equation}
\begin{split}
	f_{\chi}(q,x)=&f^{\eq}_{\chi}(q,x)\frac{Y_{\sbotb}}{Y_{\sbotb}^{\eq}}-\int_{x_0}^{x} \frac{\diff \left(f^{\eq}_{\chi}(q,y)  Y_{\sbotb}(y)/Y_{\sbotb}^{\eq}(y)\right) }{\diff y} \\& \times \exp\left( -\int_{y}^{x} \frac{\widetilde{C}(q,z)}{z H(z)}\diff z \right) \diff y. \label{eq:nonkineqsol}
\end{split}
\end{equation}

\subsection{Simplifications for the numerical solution}

In order to trace the evolution of the number density 
\begin{equation}
	n_{\chi}(x)=\frac{4\pi  g_{\chi} m_{\chi}^3}{x^3}\int\frac{q^2}{(2\pi)^3}f_\chi(q,x)\diff q\,,
\end{equation}
the phase space density $f_\chi(q,x)$ has to be computed for a large 
number of momentum modes $q$ and temperature
parameters $x$ providing a sufficiently accurate numerical approximation. 
This is a computationally expensive task. In the following we therefore
introduce analytic simplifications of the collision operators.

For the two-body decay we can find the analytic result for the collision operator, neglecting the $b$-quark mass,
\begin{equation}
	\widetilde{C}_{12}=\frac{2 g_{\sbotb}g_b |M|^2 T}{16\pi p_{\chi} E_{\chi}}\left(e^{-p_{\min}/T}-e^{-p_{\max}/T}\right)
	\label{eq:decayC}\,,
\end{equation}
where
\begin{eqnarray}
	|M|^2&=&\frac{\lambda_{\chi}^2}{g_{\chi}g_b}(m_{\sbotb}^2-m_{\chi}^2)\,,\\
	p_{\min/\max}&=&\frac{m_{\sbotb}^2-m_{\chi}^2}{2m_{\chi}^2}(E_{\chi}\mp p_{\chi})
\end{eqnarray}
and $g_{\chi}=2$, $g_b=6$, $g_{\sbotb}=3$. 
The factor of $2$ in the numerator of Eq.~\eqref{eq:decayC} accounts for the two different processes when considering the mediator and its anti-particle, which is not included in $g_{\sbotb}$.

For the inelastic scattering we consider the process $\chi b \leftrightarrow \sbot g$. 
Neglecting, again, the bottom mass we can express the collision operator as
\begin{equation}
	\label{eq:Cscatt}
	\begin{split}
	 \widetilde C_{22} = \;\,&  \frac{2 g_qT}{16\pi^2 p_\chi E_\chi} \int_{m_{\sbot}^2}^\infty  (s-m_\chi^2)\, \sigma(s)\\
	 &  \times \left(e^{-p_{\min}(s)/T}-e^{-p_{\max}(s)/T}\right) \diff s
	 \end{split}
\end{equation}
where 
\begin{equation}
  \sigma(s) = \frac{1}{4 \,| p_\chi \!\cdot p_q | }\int \!\diff\Pi_{\sbot} \diff \Pi_g (2\pi)^4\delta(p_\chi+p_q-p_{\sbot}-q_g)|{\cal M}|^2
\end{equation}
and 
\begin{equation}
  p_{\min/\max}(s) = \frac{s-m_\chi^2}{2m_\chi^2}(E_\chi\mp p_\chi)\,.
\end{equation}
Note that while $\widetilde{C}_{12}$ is an analytic function, $\widetilde{C}_{22}$
has to be numerically evaluated. We choose to precompute $\widetilde{C}_{22}$
on a two-dimensional grid in $x$ and $q=p_\chi/T$ and use an interpolation for 
the numerical evaluation of Eq.~\eqref{eq:nonkineqsol}.
We provide most of the discussion for the case when taking into 
account the decay term only, which is expected to capture the main effects. We comment
on the effect of scatterings at the end of this section.

\begin{figure}[t]
\centering
\setlength{\unitlength}{1\textwidth}
\begin{picture}(0.5,0.435)
  \put(0.006,-0.008){\includegraphics[width=0.48\textwidth]{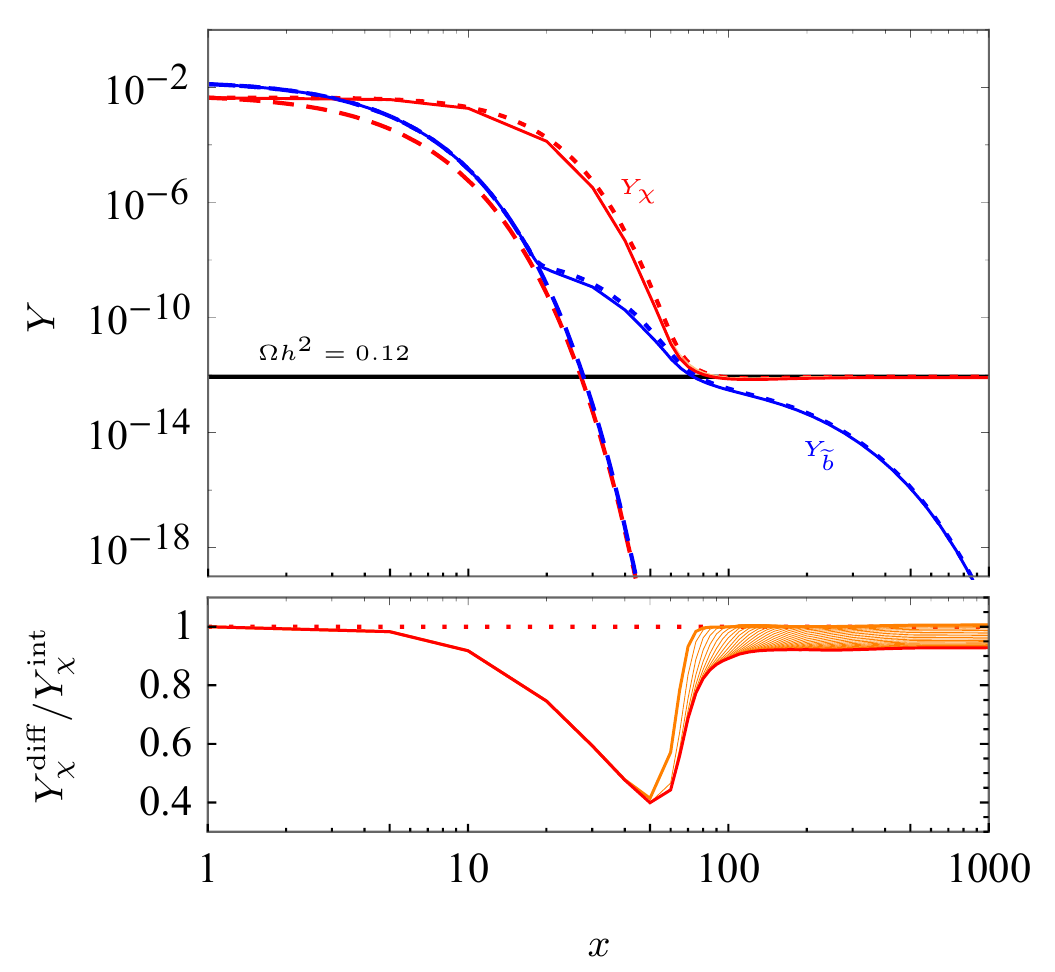}}
\end{picture}
\caption{Upper panel: Evolution for the resulting abundance of $\s b$ (blue) and $\chi$ (red) of the differential (solid) and
integrated (dotted) Boltzmann equation. The dashed curves denote the equilibrium abundances.
Lower panel: Ratio of the two abundances for $\chi$. The red solid line shows the converged result while
the orange thick and thin curves denote the first and the following iterations, respectively. Only the decay term is considered. 
}
\label{fig:diffY1}
\end{figure}

\subsection{Iterative solution of the coupled system}

The solution Eq.~\eqref{eq:nonkineqsol} of the Boltzmann equation for $f_\chi(q,x)$ requires 
as an input the evolution of the mediator abundance, $Y_{\sbot}(x)$. The latter can be obtained by 
solving the corresponding integrated Boltzmann equation, which in turn  involves $Y_\chi(x)$, that is
determined by integrating $f_\chi(q,x)$ over all momentum modes. Therefore the equations for
$f_\chi(q,x)$ and $Y_{\sbot}(x)$ form a coupled set of equations.

Here we solve this coupled set of differential equations in an iterative process. 
We start with an initial ``guess'' for $Y_{\sbot}(x)$, which we take to be the  
solution when assuming kinetic equilibrium (see below for a discussion of different choices). 
We then solve for $f_\chi(q, x)$ on a momentum-grid, and numerically compute $Y_\chi(x)$ 
using Eq.~\eqref{eq:nonkineqsol} as described in the last subsection. With this solution for 
$Y_\chi(x)$ we recalculate $Y_{\sbot}(x)$ using the integrated Boltzmann equation. We subsequently 
iterate between solving for $f_\chi(q, x)$ and $Y_{\sbot}(x)$, until we encounter sufficient convergence. 
In order to solve the differential Boltzmann equation in an acceptable time, we neglect the bottom mass 
and choose $\heff$ and $\geff$ to be evaluated at $x=50$ and constant for all times. We do not expect 
a strong dependence on these simplifications.

\begin{figure}[b]
\centering
\setlength{\unitlength}{1\textwidth}
\begin{picture}(0.5,0.63)
  \put(0.024,-0.01){\includegraphics[width=0.44\textwidth]{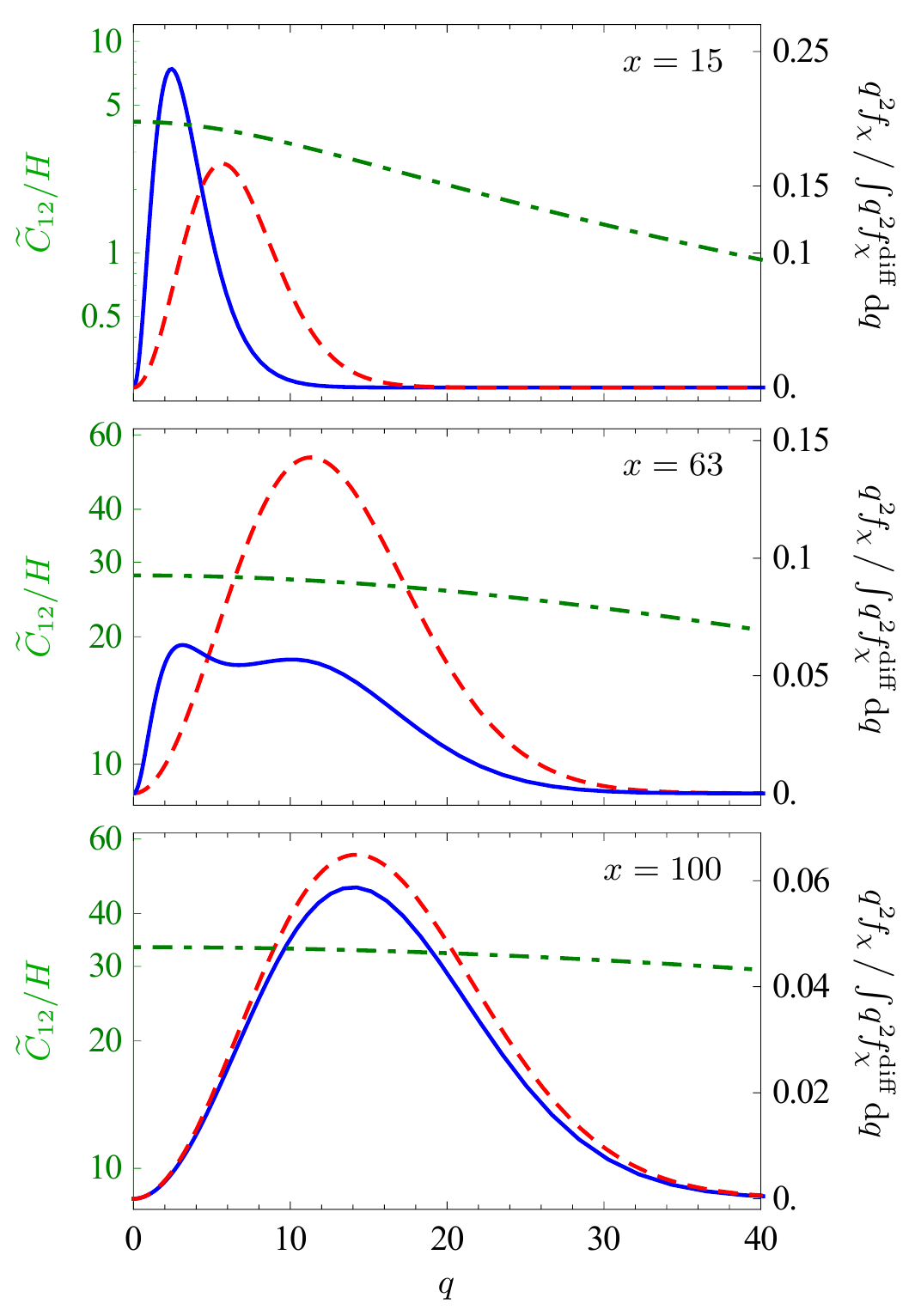}}
\end{picture}
\caption{Collision operator (normalized by the Hubble rate, green dot-dashed curves) and 
the phase space distribution of the differential (blue solid) and integrated (red dashed)
solution as a function of the momentum mode $q$ for three different times, $x=15$, 63 and 100.
The phase-space distribution is normalized to the integral over the differential solution,
$q^2 f_\chi\,/\int q^2 f_\chi^\text{diff}\,\diff q $. Only the decay term is considered.
}
\label{fig:qdep}
\end{figure}

\begin{figure*}[t]
\centering
\setlength{\unitlength}{1\textwidth}
\begin{picture}(0.95,0.303)
  \put(0.005,-0.008){\includegraphics[width=0.95\textwidth]{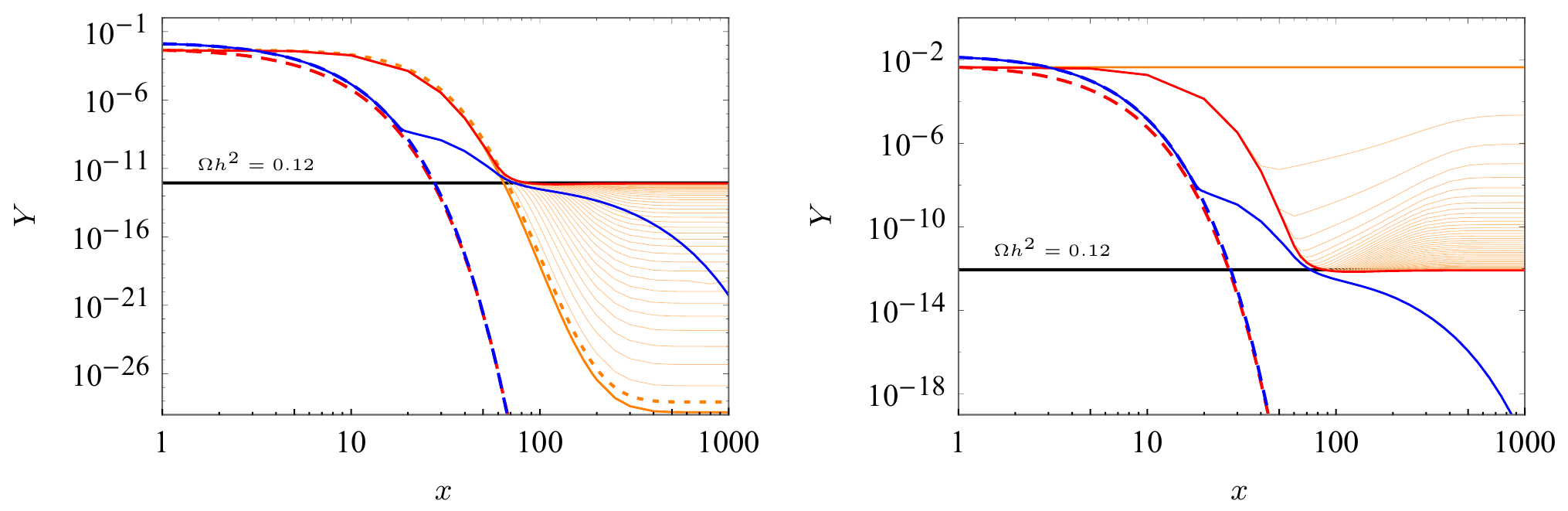}}
\end{picture}
\caption{
Evolution of the abundance of $\s b$ (blue solid) and $\chi$ (red solid) 
for two different choices of the starting point of the iteration, shown in the two panels, respectively. 
Left panel: Initial mediator abundance set to the equilibrium abundance, $Y_{\sbot}(x)=Y_{\sbot}^\eq$.
The thick and thin orange solid curves denote the first and the following iterations, respectively.
The orange dotted curve shows the integrated solution obtained for $Y_{\sbot}(x)=Y_{\sbot}^\eq$.
Right panel: Initial $\chi$ abundance set to the equilibrium abundance at relativistic temperatures, $Y_{\chi}(x)=Y_{\chi}^\eq(x\lesssim1)$.
The thick and thin orange solid curves denote the initial abundance and the following iterations, respectively.
Only the decay term is considered. As in Fig.~\ref{fig:diffY1} the dashed curves denote the equilibrium abundances.
}
\label{fig:diffY2}
\end{figure*}

The resulting evolution of the abundance $Y_\chi(x)$ for the benchmark point $m_\chi=500\,$GeV, $\msbot=510\,$GeV
is shown in Fig.~\ref{fig:diffY1} (upper panel) as a red solid curve. We compare the result to the
solution of the coupled integrated Boltzmann equation (red dotted curve) obtained under the same approximations. 
We adjust the coupling $\lambda_\chi=4.03\times10^{-7}$ so as to obtain the measured DM relic density 
for the solution of the coupled integrated Boltzmann equation.
The lower panel of Fig.~\ref{fig:diffY1} shows the ratio of the differential and integrated
solutions for $Y_\chi(x)$. While the dark matter abundance differs by up to a factor of two
at intermediate times, the final relic abundance agrees well with the corresponding
result when assuming kinetic equilibrium, with deviations below the 10\% level.

The main reason is that, for the process and the kinematical situation that is relevant here,
the collision term does not depend strongly on the momentum mode, see Fig.~\ref{fig:qdep} (dot-dashed lines).
In the same figure, we also show the result for $f_\chi(q, x)$ at various times $x$ (blue lines),
which indeed differs from an equilibrium distribution (indicated by the red dashed lines)
at intermediate times (upper and middle panel in Fig.~\ref{fig:qdep}). Nevertheless, 
around the time when the dark matter abundance freezes out, the
remaining decays of thermalized $\sbot$ tend to restore an equilibrium distribution (lowest panel).

It is interesting to observe that the total abundance obtained from the unintegrated 
Boltzmann equation is slightly \emph{below}
the result when assuming kinetic equilibrium.
This can also be understood from Fig.~\ref{fig:qdep}. 
For high temperatures, the momentum modes obtained from the differential 
solution essentially change only due to redshift.
In contrast, the kinetic equilibrium distribution populates somewhat higher modes. 
By the time the conversion gets efficient, the collision term is larger for smaller 
momentum modes. Therefore, the conversion into $\sbot$s is stronger for the 
differential solution, rendering a slightly \emph{smaller} abundance.

Let us now discuss the impact of the initial ``guess'' for $Y_{\sbot}(x)$ used for the iterative solution.
We check that the converged result is independent of the starting point of the iteration by using two 
rather different initial abundances. First, we use the equilibrium abundance $Y_{\sbot}^{\eq}(x)$ as a starting point. 
The results are shown in the left panel of Fig.~\ref{fig:diffY2}. 
The evolution of $Y_\chi(x)$ obtained in the first iteration step is shown by the thick orange line, 
and the successive iterations are indicated by the thin orange lines. The final, converged result 
(thick red line) agrees well with the result obtained in Fig.\,\ref{fig:diffY1}. The same is true for 
$Y_{\sbot}(x)$ (solid blue line). On the other hand, we would like to point out that the first iteration and the
converged result are rather far apart. This means that it is crucial to solve for the coupled set of equations,
allowing for deviations $Y_{\sbot}(x)\not= Y_{\sbot}^{\eq}(x)$. For curiosity, we note that if one would compare the
differential with the integrated result for $Y_\chi(x)$ while fixing $Y_{\sbot}(x)= Y_{\sbot}^{\eq}(x)$, one
would find an ${\cal O}(10)$ difference in the final abundance
(see orange dotted versus solid line in  the left panel of Fig.~\ref{fig:diffY2}), while the corresponding difference
for the converged results is below $\sim 10\%$. Hence, the partial freeze-out
of the mediator $\sbot$ and its subsequent decay into $\chi$ are crucial for the conclusion that the
impact of deviations from kinetic equilibrium on the relic density is small.

\begin{figure}[t]
\centering
\setlength{\unitlength}{1\textwidth}
\begin{picture}(0.47,0.322)
 \put(0.0,-0.008){\includegraphics[width=0.47\textwidth]{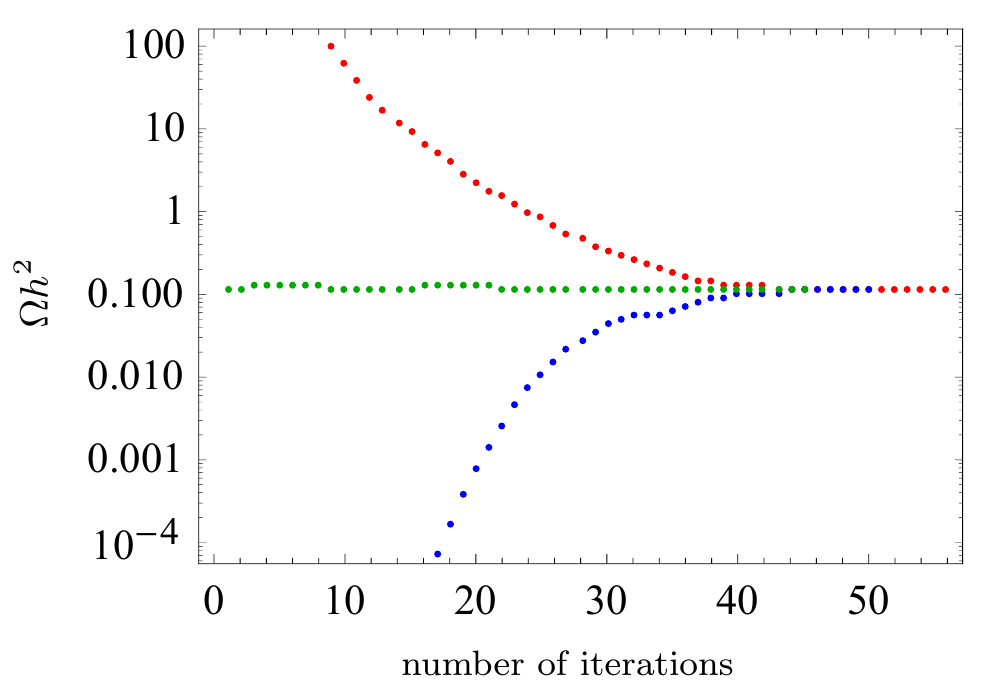}}
\end{picture}
\caption{Relic density obtained from the iterative solution as a function of
the iteration step for the three different initial abundances discussed here: $Y_{\chi}(x)=Y_{\chi}^\eq(x\lesssim1)$
(red points, converging from above), $Y_{\sbot}(x)=Y_{\sbot}^\eq$ (blue points, converging
from below) and $Y_{\sbot}(x)$ set to the solution of the coupled integrated Boltzmann equation
(green points, starting in the middle).
}
\label{fig:itstep}
\end{figure}

Second, we consider an extreme possibility and initially set $Y_{\chi}(x)$ to be constant and equal to the relativistic equilibrium density.
In this case we start the iteration with the computation of $Y_{\sbot}(x)$.
The resulting iterative solutions for $Y_{\chi}(x)$ are shown in right panel of Fig.~\ref{fig:diffY2} (orange lines).
Again, the converged result for $Y_{\chi}(x)$ (red solid line) and $Y_{\sbot}(x)$ (solid blue line)
agree well with those shown in Fig.\,\ref{fig:diffY1}.

The convergence of the final relic density for the three different choices of initial abundances is shown in Fig.~\ref{fig:itstep}. 
Indeed, the converged results agree, indicating that the iterative scheme is stable and leads to a unique result.

Next we want to check if the situation changes drastically when including also $2\rightarrow 2$ scattering processes. 
Due to the increase in numerical complexity described above, we consider the leading process 
$ \chi b \leftrightarrow \sbot g$ expected to capture the main effects.
In order to estimate the physical contributions from hard scatterings, we perform regularizations on the level of the scattering cross section by 
introducing a cut-off $s_{\min}=(\msbot+1\,\text{GeV})^2$ and additionally a regulator at the matrix 
element level of $1/t^2\rightarrow 1/(t^2+(1\,\text{GeV})^4)$. 
In addition, we restrict the integration over the angle $\theta_t$ between $b$ and $g$ in the center-of-mass frame
to $\cos\theta_t \in [-0.9,0.9]$. 

Again, we solve the coupled system in an iterative approach as described above, but taking scatterings into account.
As before, we then compare the converged result for the final relic density with the corresponding
result obtained when assuming kinetic equilibrium.
We find that the relative deviations in the resulting 
relic density 
stay below $10\%$. Furthermore, the deviation for $Y_{\chi}(x)$ for intermediate
times becomes smaller. This is expected, because scatterings increase the conversion rates at smaller~$x$.

Altogether, we find that the impact of deviations from kinetic equilibrium on the final relic abundance
is rather mild, below the $10\%$ level. This justifies using integrated Boltzmann equations for $Y_{\chi}(x)$ 
and $Y_{\sbot}(x)$.

\end{appendix}

\bibliography{bibliography}{}

\end{document}